\documentclass[sigconf]{acmart}
 
\def\BibTeX{{\rm B\kern-.05em{\sc i\kern-.025em b}\kern-.08emT\kern-.1667em\lower.7ex\hbox{E}\kern-.125emX}}
      
\usepackage{multirow}

\usepackage{amsmath,amssymb,amsfonts}
\usepackage{graphicx}
\usepackage{textcomp}

\usepackage[ruled,vlined]{algorithm2e}
\usepackage{algpseudocode}

\usepackage{xcolor}
\usepackage{braket}

\usepackage[normalem]{ulem} 

\newcommand{\Hop}{H}
\newcommand{\para}{\vec{\theta}}
\newcommand{\boldx}{\bold{x}}
\newcommand{\boldy}{\bold{y}}
\newcommand{\Eloc}{E_{loc}}
\newcommand{\aop}{\hat{a}}
\newcommand{\adop}{\hat{a}^{\dagger}}
\newcommand{\im}{{\rm i}}
\newcommand{\samples}{\mathcal{S}}
\newcommand{\real}{{\rm Re}}

\newcommand{\uniquesample}{N_u}

\newcommand{\wyj}[1]{{{#1}}}

\floatname{algorithm}{Algorithm}

\copyrightyear{2023}
\acmYear{2023}
\setcopyright{acmlicensed}\acmConference[SC '23]{The International Conference for High Performance Computing, Networking, Storage and Analysis}{November 12--17, 2023}{Denver, CO, USA}
\acmBooktitle{The International Conference for High Performance Computing, Networking, Storage and Analysis (SC '23), November 12--17, 2023, Denver, CO, USA}
\acmPrice{15.00}
\acmDOI{10.1145/3581784.3607061}
\acmISBN{979-8-4007-0109-2/23/11}

\begin{document}

\title{NNQS-Transformer: an Efficient and Scalable Neural Network \\ Quantum States Approach for \textit{Ab initio} Quantum Chemistry}

\author{Yangjun Wu}
\affiliation{
  \institution{Institute of Computing Technology, Chinese Academy of Sciences, UCAS}
  \city{Beijing}
  \country{China}}
\email{wuyangjun21s@ict.ac.cn}

\author{Chu Guo}
\authornote{Corresponding authors}
\affiliation{
  \institution{ Hunan Normal University}
  \city{Changsha}
  \country{China}}
\email{guochu604b@gmail.com}

\author{Yi Fan}
\affiliation{%
  \institution{University of Science and Technology of China}
  \city{Hefei}
  \country{China}}
\email{fanyi@mail.ustc.edu.cn}

\author{Pengyu Zhou}
\affiliation{
  \institution{Institute of Computing Technology, Chinese Academy of Sciences, UCAS}
  \city{Beijing}
  \country{China}}
\email{zhoupengyu19@mails.ucas.ac.cn}

\author{Honghui Shang}
\authornotemark[1]
\affiliation{%
  \institution{University of Science and Technology of China}
  \city{Hefei}
  \country{China}}
\email{shanghui.ustc@gmail.com}

\begin{abstract}

Neural network quantum state (NNQS) has emerged as a promising candidate for quantum many-body problems, but its practical applications are often hindered by the high cost of sampling and local energy calculation. 
We develop a high-performance NNQS method for \textit{ab initio} electronic structure calculations. The major innovations include: 
(1) A transformer based architecture as the quantum wave function ansatz; 
(2) A data-centric parallelization scheme for the variational Monte Carlo (VMC) algorithm which preserves data locality and well adapts for different computing architectures; 
(3) A parallel batch sampling strategy which reduces the sampling cost and achieves good load balance; 
(4) A parallel local energy evaluation scheme which is both memory and computationally efficient; 
(5) Study of real chemical systems demonstrates both the superior accuracy of our method compared to state-of-the-art and the strong and weak scalability for large molecular systems with up to $120$ spin orbitals.
 \end{abstract}


\maketitle

\section{Introduction}

Electronic structure calculation based on first-principle quantum mechanics is an elementary tool for predicting the chemical and physical properties of matter.
The grand challenge of \textit{ab initio} electronic structure calculations is to solve the Schr$\ddot{\text{o}}$dinger equation $\Hop\psi = E\psi$ of an interacting many-body system of $n$ electrons and $m$ nuclei, whose Hamiltonian $\Hop$ can be written as (under the Born-Oppenheimer approximation to freeze the nuclear degrees of freedom),
\begin{align}\label{eq:ham}
	\Hop =   -\displaystyle\sum_{i=1}^{n}\frac{1}{2}\nabla_{i}^{2} - \displaystyle\sum_{i=1}^{n}\displaystyle\sum_{A=1}^{m}\frac{Z_A}{r_{iA}} + \displaystyle\sum_{i=1}^{n}\displaystyle\sum_{j>i}^{n}\frac{1}{r_{ij}}, 
\end{align}
where $r_{ij} = |r_i - r_j|$ denotes the distance between two electrons at positions $i$ and $j$, $r_{iA}$ denotes the distance between an electron at position $i$ and a nucleus at position $A$, and $Z_A$ is the atomic number of the nucleus at position $A$. The state of interest is the ground state $\vert \psi\rangle$ that minimizes the energy: $E = \langle \psi\vert \Hop \vert \psi\rangle $.

However, the exact solution for the ground state is generally impossible since 
the dimension of the state space for wave functions grows exponentially with the number of electrons. As a result, the full configuration interaction (FCI) method, which takes into account all the possible states, is currently limited within $24$ electrons and $24$ orbitals even with the best supercomputer~\cite{VogiatzisJong2017}.
In order to deal with this problem, various approximation techniques have been developed, such as the Hartree-Fock~(HF) theory and density functional theory~(DFT),  which are able to scale up to thousands (or even more) of electrons by reducing the problem to a single-electron problem instead, but they could introduce errors in the calculation of physical properties of the system, and may be inaccurate for certain types of systems.
The truncated configuration interaction~(CI)~\cite{DAVIDSHERRILL1999} considers only excitations above the HF reference state up to a fixed order, while the coupled cluster (CC) methods consider certain nonlinear combinations of the excitation operators up to a fixed order. In particular, the CC methods could often give reasonable solutions with a limited computational cost, thus referred to as the gold standard. However, they could still fail in presence of strong static correlations~\cite{BulikScuseria2015}. The density matrix renormalization group (DMRG) method~\cite{White1992,White1993} could give solutions with comparable precision to FCI and has been used to study molecular systems with up to $30$ electrons ($108$ orbitals)~\cite{BrabecVeis2020,LarssonChan2022}, however it is mainly limited by the expressivity of the underlying matrix product state ansatz which can only efficiently parameterize finitely correlated states~\cite{GarciaCirac2007}. DMRG is also very unfriendly for large-scale parallelization.
The conventional quantum Monte Carlo (QMC) methods~\cite{FoulkesRajagopal2001,NeedsRios2010,AustinLester2012}, including the variational Monte Carlo (VMC) and diffusion Monte Carlo (DMC), can directly deal with real space calculations by using an anti-symmetric wave function ansatz which accounts for the Pauli exclusion principle of electrons (the first quantized formalism). However high precision calculations using first quantized QMC methods in real space are still quite limited due to the high computational cost. Prominent approaches in this direction include the FermiNet~\cite{PfauFoulkes2020} and PauliNet~\cite{HermannNoe2020} which are currently limited within $30$ electrons.

In 2017, Carleo and Troyer propose the neural network quantum state (NNQS) algorithm for many-spin systems, which parameterizes the wave function as a neural network and uses the VMC algorithm to optimize the parameters~\cite{CarleoTroyer2017}. In particular they demonstrate that using a very simple neural network ansatz, namely the restricted Boltzmann machine (RBM) that is essentially a dense layer plus a nonlinear activation, could already achieve competitive accuracy to existing tensor network algorithms. Till now the performance of NNQS has been demonstrated in a wide variety of many-spin problems~\cite{CarleoTroyer2017,ChooCarleo2019,SharirShashua2020,SchmittHeyl2020,YuanDeng2021,ZhaoLiang2022} as well as many-fermion problems~\cite{MorenoStokes2022}. 
NNQS is a promising alternative for quantum many-body problems since 1) there is no fundamental limit on the expressive power of (deep) neural network ansatz for many-body quantum state~\cite{DengSarma2017,GlasserCirac2018,SharirCarleo2022,GaoDuan2017}; 2) the computational cost scales polynomially in general and 3) it often allows transparent large-scale parallelization as a general feature of Monte Carlo methods.
In 2020, RBM was first introduced for electronic structure calculations by converting the fermionic problem in the second quantized formalism into a many-spin problem using Jordan-Wigner (JW) transformation~\cite{ChooCarleo2020}, where it is demonstrated that RBM could achieve higher accuracy than CC with excitations up to second order (CCSD) and third order (CCSD(T)) for a wide variety of molecular systems within $24$ spin orbitals (qubits). 
Two later works pushed the simulation scale to $30$ qubits~\cite{BarrettLvovsky2022} and $76$ qubits~\cite{Zhao_2023} respectively by using autoregressive neural network ansatz to reduce the sampling cost.

Despite the polynomial scaling of the computational cost of NNQS, for electronic structure calculations, it is still extremely difficult to scale up the application size to $100$ qubits or more (while maintaining comparable or higher accuracy than CC or DMRG) up to date. The challenge is mainly threefold: 1) For conventional neural networks such as RBM or the convolutional neural network (CNN), a Markov chain (MC) sampling algorithm is used to generate samples for energy and gradient estimation, which could be very inefficient in certain situations with low acception rate. Moreover, for these neural networks one also has to use the stochastic reconfiguration (SR) technique~\cite{SorellaCapriotti2000,SorellaRocca2007} for stable convergence to global minimum, for which one needs to (approximately) compute the inverse of the $M\times M$ SR matrix  for a neural network with $M$ parameters, thus greatly prohibiting the usage of very deep neural networks as well as the scalability to a large number of processes. As an example a recent application of NNQS on the new Sunway supercomputer used a CNN with only around $50,000$ parameters~\cite{ZhaoLiang2022}; 2) 
The chemical bonding in molecular systems can be very different in nature and of very different strengths, while for practical applications it is important to use a model which is easily extensible to deep architectures for better expressivity and thus can be versatilely used for different molecular systems; 
3) For molecular systems in the second quantized formalism, the number of Pauli strings after the JW transformation often scales as $O(N^4)$ for $N$ qubits, therefore in large applications, evaluation of the local energy in VMC could be very time-consuming. For example 3) is the major limiting factor for Ref.~\cite{BarrettLvovsky2022} to go beyond $30$ qubits.

In view of these challenges, we believe a highly efficient and scalable NNQS method is in need, which is an important step towards large-scale applications in electronic structure calculations and also acts as a powerful tool for cross validation with existing methods.

The major innovations of this work can be summarized as follows.
\begin{itemize}
\item  A deep neural network wave function ansatz \wyj{(QiankunNet~\cite{shang2023solving})} is used for VMC optimization, which consists of a transformer made of stacked decoders for the amplitude and a multilevel perceptron (MLP) for the phase, inspired by the enormous success of transformer based architectures in natural language processing (NLP) and other classical machine learning tasks~\cite{Transformer2017,radford2018gpt,radford2019gpt2}.
  
\item A data-centric parallelization scheme based on MPI communication for VMC algorithms using autoregressive wave function ansatz, which maximally preserves the data locality and well adapts to different computing architectures. The total amount of data communication in this scheme only scales as $O(\uniquesample)$ for $\uniquesample$ the total number of unique samples and $O(M)$ for $M$ parameters.

\item  A parallel batch sampling strategy which makes use of the autoregressive property of our wave function ansatz and achieves low sampling cost and good load balance.

\item A local energy calculation scheme parallelized over batches of samples, powered by a highly compressed data structure for the Hamiltonian together with a fused design of nonzero Hamiltonian entry evaluation and local energy calculation to reduce memory cost, and a sample-aware evaluation scheme together with an efficient storage of the unique samples as a lookup table to reduce the computational cost.

\item Study of real chemical systems shows that our method could reach superior accuracy compared to state-of-the-art NNQS methods, and achieve both strong and weak scaling for the benzene molecular system with $120$ qubits.

\end{itemize}
Our parallelized implementation of the transformer based NNQS can be easily transplanted to different high-performance computing architectures as long as the pytorch library~\cite{pytorch} is supported and the local energy evaluation function is implemented on each process. Our work opens a verge of applying state-of-the-art transformer based architectures for large-scale electronic structure calculations.

\section{Background}
In this section, we review the basic ideas of the VMC algorithm, autoregressive neural networks, typical quantum chemistry Hamiltonians, as well as state-of-the-art NNQS approaches for electronic structure calculations.

\subsection{Variational Monte Carlo algorithm}

\begin{figure}
	\centerline{\includegraphics[width=0.95\columnwidth]{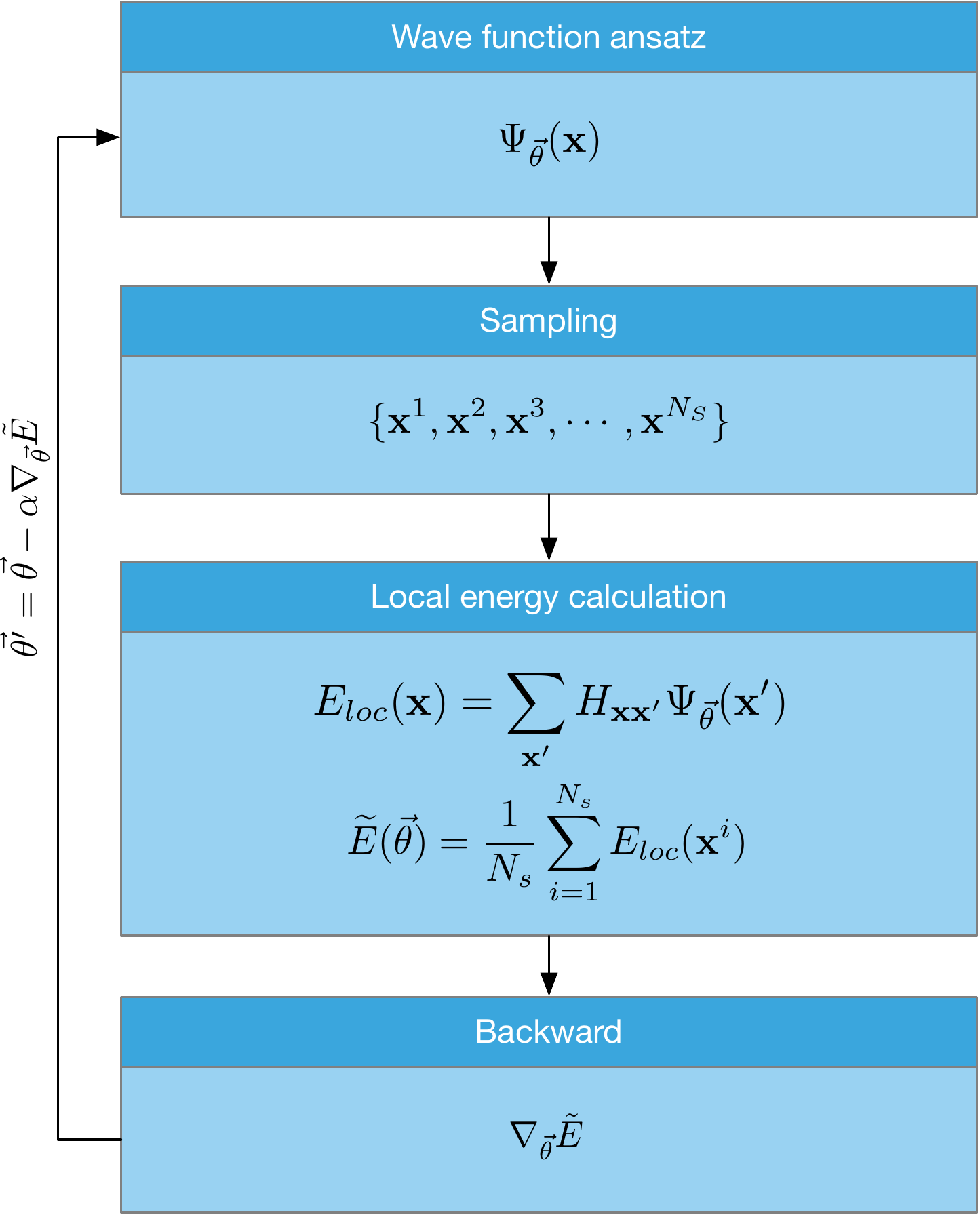}}
	\caption{Schematic of the workflow for the serial NNQS calculation.}
	\label{fig:serial_flow}
\end{figure}

Assuming that one has chosen a variational ansatz for the ground state denoted as $\vert \psi_{\para}\rangle$, with $\para$ the parameters to be optimized, then the energy of the system can be written as a function of $\para$:
\begin{align}\label{eq:energy1}
E(\para) = \frac{\langle \psi_{\para} \vert \Hop\vert \psi_{\para}\rangle}{\langle \psi_{\para}\vert \psi_{\para}\rangle },
\end{align}
where the denominator indicates that the ansatz $\vert \psi_{\para}\rangle$ may not be normalized in general. The ground state can be approached by minimizing $E(\para)$ against $\para$. Expanding Eq.(\ref{eq:energy1}) in the computational basis, we obtain
\begin{align}\label{eq:energy2}
E(\para) =& \frac{ \sum_{\boldx, \boldx'} \langle \psi_{\para} \vert \boldx \rangle\langle \boldx \vert \Hop \vert \boldx'\rangle\langle\boldx' \vert \psi_{\para}\rangle   }{ \sum_{\boldy} \langle \psi_{\para}\vert \boldy\rangle\langle\boldy\vert  \psi_{\para}\rangle } \nonumber \\
=& \frac{ \sum_{\boldx} \left(\sum_{\boldx'}H_{\boldx\boldx'} \psi_{\para}(\boldx') / \psi_{\para}(\boldx) \right) |\psi_{\para}(\boldx)|^2 }{\sum_{\boldy}|\psi_{\para}(\boldy)|^2 },
\end{align}
where $\boldx$, $\boldx'$, $\boldy$ denote specific bitstrings. We have written $H_{\boldx \boldx'} = \langle \boldx\vert \Hop\vert \boldx'\rangle$ and $ \psi_{\para}(\boldx) = \langle \boldx\vert \psi_{\para}\rangle $ as the probability amplitude of the wave function ansatz $\vert \psi_{\para}\rangle$ in basis $\vert \boldx\rangle$ in the second line of Eq.(\ref{eq:energy2}). Defining the local energy $\Eloc(\boldx)$ as
\begin{align} \label{eq:Eloc}
\Eloc(\boldx) = \sum_{\boldx'}H_{\boldx\boldx'} \psi_{\para}(\boldx') / \psi_{\para}(\boldx), 
\end{align}
and the (unnormalized) probability $p_{\para}(\boldx) = |\psi_{\para}(\boldx)|^2$, Eq.(\ref{eq:energy2}) can be rewritten as
\begin{align}\label{eq:energy3}
E(\para) = \frac{\sum_{\boldx}\Eloc(\boldx) p_{\para}(\boldx) }{\sum_{\boldy}p_{\para}(\boldy)} = \mathbb{E}_{p}[\Eloc(\boldx)].
\end{align}
It is impossible to exactly evaluate Eq.(\ref{eq:energy3}) in general as there exists an exponential number of different bitstrings. However, one could approximately evaluate Eq.(\ref{eq:energy3}) by sampling from the probability distribution $p_{\para}(\boldx)$ to obtain a set of $N_s$ samples, denoted as $\{\boldx^1, \boldx^2, \dots, \boldx^{N_s}\}$, and then averaging over them
\begin{align}\label{eq:energy}
\tilde{E}(\para) = \frac{1}{N_s} \sum_{i=1}^{N_s} \Eloc(\boldx^i),
\end{align}
where we have used $\tilde{E}(\para)$ instead of $E(\para)$ to stress that the former is only an approximation of the latter. To this end, we can see that as long as one can efficiently sample from $p_{\para}(\boldx)$ (which is the case if we can evaluate $\psi_{\para}(\boldx)$ efficiently for each $\boldx$), and efficiently find those nonzero entry $\Hop_{\boldx \boldx'}$ together with $\boldx'$, then Eq.(\ref{eq:energy}) can be efficiently evaluated. To speed up the calculation, it is often advantageous to use a gradient-based optimizer instead of gradient-free ones. With the obtained samples, one could compute the approximate gradient of Eq.(\ref{eq:energy3}) using automatic differentiation~\cite{BarrettLvovsky2022}:
\begin{align}
\nabla_{\para}\tilde{E} = 2\real\left(\mathbb{E}_p \left[\left(\Eloc(\boldx) - \mathbb{E}_p\left[\Eloc(\boldx) \right] \right) \nabla_{\para} \ln \left(\Psi_{\para}^{\ast}(\boldx)\right) \right] \right).
\end{align}
Similarly, $\nabla_{\para}\tilde{E}$ is used to approximate the exact gradient $\nabla_{\para}E$. After that, the parameters $\para$ can be updated based on $\nabla_{\para}\tilde{E}$ and the optimizer, finishing one iteration of the VMC algorithm.
The workflow of the serial NNQS algorithm is demonstrated in Fig.~\ref{fig:serial_flow}.


\subsection{Autoregressive neural networks}
A high-dimension probability distribution $\pi(\boldx)$ is said to be autoregressive if it can be evaluated as follows:
\begin{align}\label{eq:auto}
\pi(\boldx) = \pi_{x_1} \pi_{x_2|x_1} \cdots \pi_{x_{N}|x_{N-1}, \dots, x_1},
\end{align}
where $N$ is the length of $\boldx$ and $x_i$ the $i$-th element of $\boldx$, $\pi_{x_i|x_{i-1}, \dots, x_1}$ denotes the conditional probability distribution of $x_i$ on the observed sequence from $x_1$ to $x_{i-1}$. The autoregressive property makes it very easy to sample from $\pi(\boldx)$: one can first sample from the single-variate probability distribution $\pi(x)$ and get an outcome $x_1$, then one can sample from another single-variate probability distribution $\pi(x|x_1)$ and get an outcome $x_2$, following this procedure one can obtain one sample $\boldx = \{x_1, \dots, x_N\}$ from $\pi(\boldx)$ by only performing $N$ local samplings from $N$ single-variate probability distributions. This sampling algorithm will be referred to as the autoregressive sampling (demonstrated in Fig.~\ref{fig:bas}(a)).

The advantage of using autoregressive neural networks based NNQS is at least two-fold: 1) It is efficient in the sense that only local samplings are required (the local state space size is only $2$ for spin systems); 2) The obtained samples are exact in the sense that they are free from sampling inaccuracies such as the auto-correlation problem of the MC sampling, and importantly, we observe in practice that we can often easily converge to the ground state without using the SR technique, thus one can use very deep neural networks with a large number of parameters and scale up the algorithm to a large number of processes.

\subsection{Quantum Chemistry Hamiltonians}
Once a finite single-electron basis set has been chosen, one can expand the continuous Hamiltonian in this basis and obtain a second quantized Hamiltonian in the following form:
\begin{align}\label{eq:eham}
\Hop^e = \sum_{p, q} h^{p}_q \adop_p\aop_q + \frac{1}{2}\sum_{p,q,r,s}g^{p,q}_{r,s}\adop_p\adop_q\aop_r\aop_s,
\end{align}
with $h^{p}_q$ and $g^{p,q}_{r,s}$ the one- and two-electron integrals, $\adop_p$ and $\aop_q$ the creation and annihilation operators. For convenience of using arbitrary neural networks, the JW transformation can be used to convert $\Hop^e$ into a many-spin Hamiltonian, which generally takes the form
\begin{align}\label{eq:sham}
\Hop^s = \sum_{i=1}^{N_h} c_i P_i,
\end{align}
where each $P_i$ is the tensor product of Pauli spin operators $\{I, X, Y, Z\}$ of length $N$, referred to as a Pauli string, and $c_i$ is a real coefficient. $N_h$ denotes the total number of Pauli strings. We note that the integrals in Eq.(\ref{eq:eham}) can be calculated using the package PySCF~\cite{pyscf} while the JW transformation can be done using the package OpenFermion~\cite{openfermion}. For quantum chemistry Hamiltonians, $N_h$ often scales as $O(N^4)$, which means that for each input bitstring $\boldx$, there could exist $O(N^4)$ $\boldx'$s with nonzero $\Hop^s_{\boldx \boldx'}$. As a result for large $N$, evaluating the local energy in Eq.(\ref{eq:Eloc}) can be very expensive and storing all the $\boldx'$s that are nontrivially coupled to $\boldx$ could use a large amount of memory. In this work, we only use the spin Hamiltonian $\Hop^s$, and we will denote it as $\Hop$ for simplicity in the following.

\subsection{Current state of the art}
Electronic structure calculation using NNQS in the second quantized formalism is still in its early stage.
The RBM was first introduced for electronic structure calculations in 2020, where the largest molecular system studied is the H$_2$O molecular system in the 6-31G basis ($26$ qubits) due to the high sampling cost~\cite{ChooCarleo2020}. The autoregressive neural network is first introduced as the wave function ansatz in NNQS to study quantum many-spin systems also in 2020~\cite{SharirShashua2020}, and a parallel implementation of autoregressive neural network for many-spin systems is demonstrated in 2021~\cite{ZhaoVeerapaneni2021}. Autoregressive neural network is first introduced for electronic structure calculations in 2022, which uses a multilevel perceptron (MLP) with hard-coded pre- and postprocessing to ensure the autoregressive property and is referred to as neural autoregressive quantum state~(NAQS)~\cite{BarrettLvovsky2022}. It is demonstrated that NAQS can reach superior accuracy for all the molecular systems studied with up to $30$ qubits compared to CC methods and RBM. A parallel implementation of an autoregressive neural network named MADE (masked autoencoder for distribution estimation~\cite{MADE}), with optimized local energy calculation on GPU, pushed the simulation scale to $76$ qubits (the CNa$_2$O$_3$ molecular system)~\cite{Zhao_2023}. However in the last work the accuracy is about the same or only mildly improved against CCSD results for all the test cases, and is worse than Ref.~\cite{BarrettLvovsky2022} for molecular systems within $30$ qubits. 

To summarize, for electronic structure calculations using NNQS~\cite{BarrettLvovsky2022} in the second quantized formalism, NAQS gives the most accurate results up to date, but the simulation scale is currently limited within $30$ qubits. MADE~\cite{Zhao_2023} with GPU optimized local energy calculation reaches the largest simulation scale ($52$ electrons, $76$ qubits), but is not as accurate as NAQS.

\section{Innovations}\label{sec:innovations}

\subsection{QiankunNet: Transformer based deep neural network as wave function ansatz}

\begin{figure}
\centerline{\includegraphics[width=0.95\columnwidth]{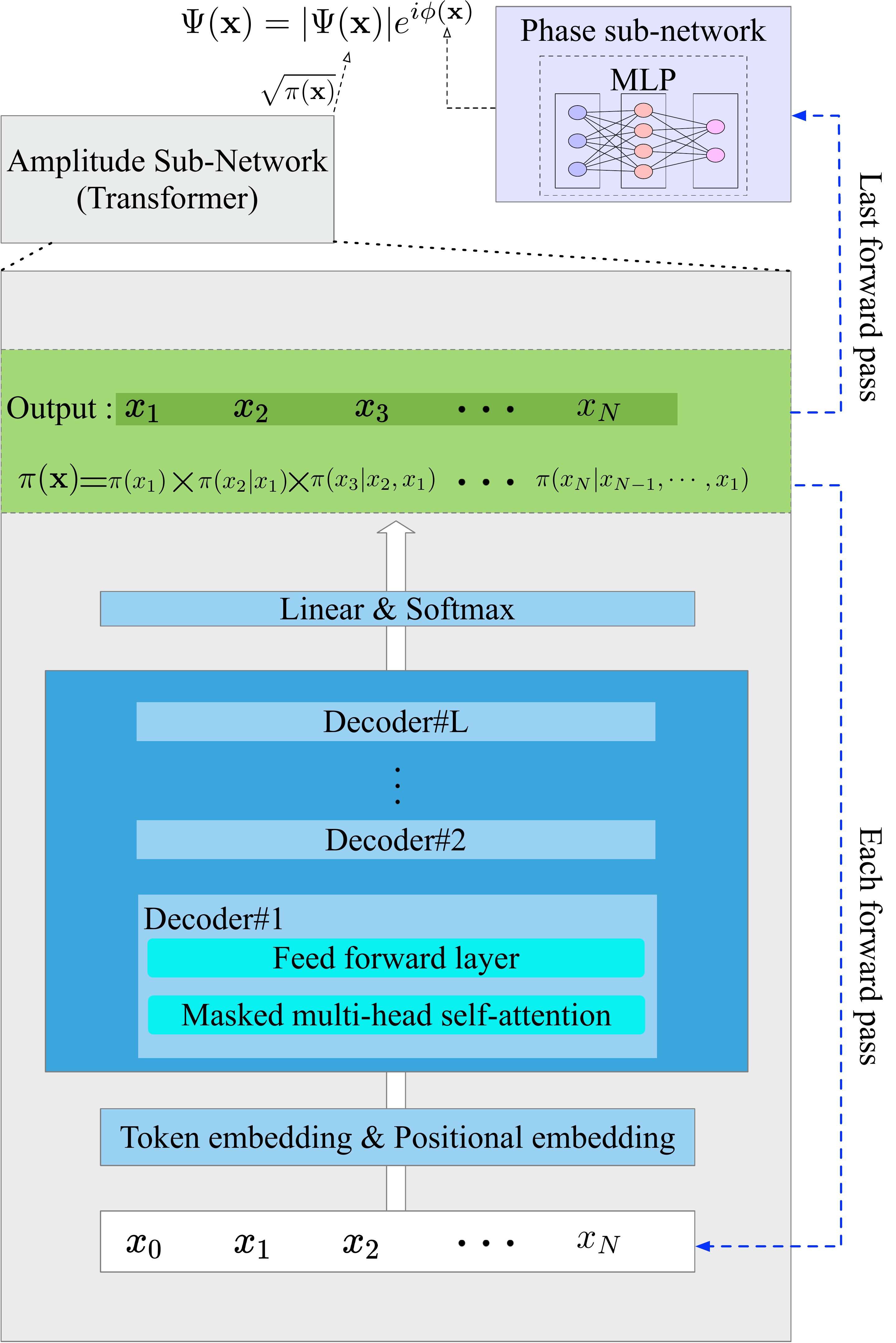}}
\caption{\wyj{QiankunNet:} The transformer based architecture for the wave function ansatz, where the transformer is used for the amplitude and the multilevel perception is used for the phase.}
\label{fig:transformer}
\end{figure}

The difficulty of solving quantum many-body problems originates from the fact that quantum states live in exponentially large Hilbert spaces ($2^{N}$ for $N$ qubits). To tackle this difficulty we borrow ideas from classical machine learning, where there also exist many problems for which the solutions live in exponentially large state spaces. As an example, in NLP a sentence could contain many words, each of which may be taken from a vocabulary of size $30522$. If one sets a maximal sentence length to be $512$, then the possible sentences constitute a state space of size $30522^{512}$~\cite{devlin2018bert}, which is larger than the sizes of most quantum many-body problems that can be numerically solved. Nevertheless, the patterns that different sentences can appear are usually limited by a few (unknown) grammatical rules, and
with state-of-the-art language modeling techniques using the transformer-based architectures (e.g. ChatGPT)~\cite{radford2018gpt,radford2019gpt2}, these rules can already be learned with high precision.

In the context of NLP, the language models are often designed such that they are generative models which are able to accept variable-length inputs and/or generate variable-length outputs, and they are also designed to be autoregressive with a natural interpretation: given a sequence of observed words, the model tries to guess the next word by sampling from the estimated probability distribution of it conditioned on the observed ones (“given the text so far, what should the next word be?”). From the measurement point of view, a quantum state is similar to a language model: one could obtain a sample (bitstring) from the quantum state by measuring each qubit of it sequentially, and in general, the $i$-th measurement outcome will be dependent on all the previous observed outcomes because of quantum correlations.

It is known that the transformer is able to learn distant correlations with a constant number of operations~\cite{Transformer2017}, while in comparison for other neural networks such as CNN the number of required operations grows linearly with the distance~\cite{JohnKremer2001}. Inspired by the enormous successes made by the transformer based architectures in NLP and other machine learning tasks, we use a customized transformer based deep neural network as the wave function ansatz. Concretely, we decompose the amplitude part and phase part of the wave function ansatz
\begin{align}
\Psi(\boldx) = |\Psi(\boldx)| e^{\im \phi(\boldx)},
\end{align}
and then we use transformer to represent the probability $|\Psi(\boldx)|^2$, and use MLP to represent the phase $\phi(\boldx)$. In comparison, Refs.~\cite{BarrettLvovsky2022,Zhao_2023} have used MLP based neural networks for both the amplitude and the phase. Our wave function ansatz is demonstrated in Fig.~\ref{fig:transformer}. In comparison with the general transformer, we have only used decoders since our problem is unsupervised. Our model can be easily made very deep by stacking a large number of decoders. In addition, our ansatz is autoregressive since only the amplitude contributes to the probability $|\Psi(\boldx)|^2$ and the amplitude part is autoregressive by design, therefore we can use very efficient sampling algorithms designed for autoregressive models.

\begin{figure}[!htbp]
	\centerline{\includegraphics[width=\columnwidth]{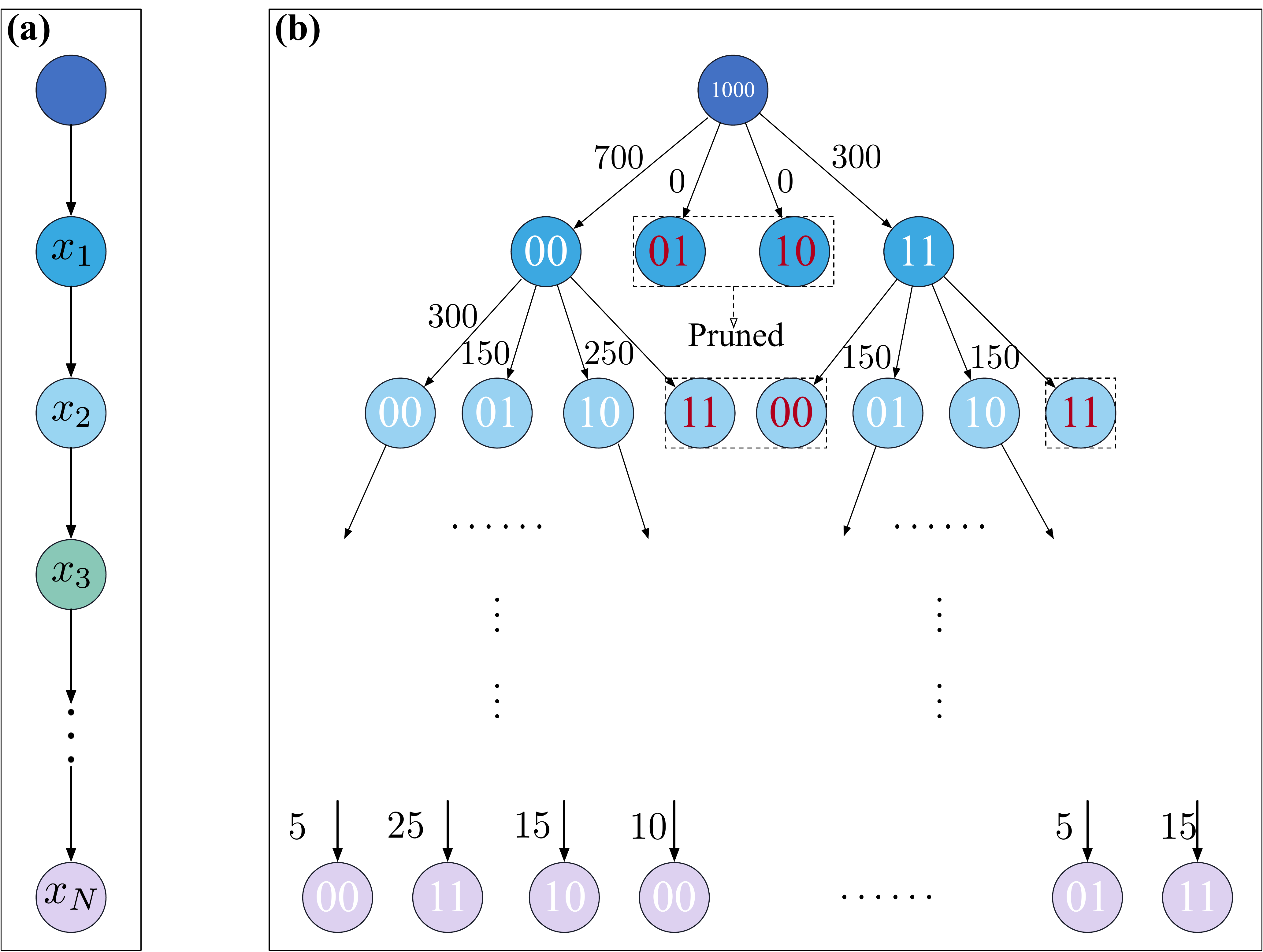}}
	\caption{(a) The autoregressive sampling algorithm which generates one sample per run. (b) The batch autoregressive sampling algorithm which generates $N_s$ samples per run. We have assumed that two qubits are sampled in each local sampling step, thus a quadtree. Each circle in (b) corresponds to a particular local sampling outcome, and the number on the edge pointing to the circle means the weight. $N_s$ can be chosen to be any number ($N_s=1000$ is used as an example). }
	\label{fig:bas}
\end{figure}

The autoregressive sampling returns one sample per run, during which $N$ local samplings are performed. 
For autoregressive wave function ansatz Ref.~\cite{BarrettLvovsky2022} proposes a batch autoregressive sampling (BAS) algorithm which is much more efficient than the autoregressive sampling. The basic idea of BAS is that in each local sampling step of the autoregressive sampling, one generates a batch of samples instead of a single one. 
Concretely, one first generates $N_s$ samples ($N_s$ could easily be as large as $10^{12}$) from the first local sampling step (namely sampling from $\pi(x)$), and only stores the unique samples with their weights. Then in the $i$-th step, one computes the probability distribution $\pi_{x|x_{i-1}, \dots, x_1}$ for each unique sample $x_{i-1}\cdots x_1$ generated in the previous step (assuming that $w_{x_{i-1}, \dots, x_1}$ is the corresponding weight), and performs local samplings from $\pi_{x|x_{i-1}, \dots, x_1}$ to obtain exactly $w_{x_{i-1}, \dots, x_1}$ samples. The intermediate samples with zero weights are pruned in each layer.
In this way, one can generate $N_s$ samples at the $N$-th step, stored as $\uniquesample$ unique samples with their weights.

The autoregressive sampling and the BAS are demonstrated in Fig.~\ref{fig:bas}. 
For transformer based architectures, calculating the conditional probability distribution $\pi(x|x_{i-1}, \dots, x_1)$ has a computational cost that roughly scales as $O(i^2)$. As a result the cost of the autoregressive sampling for $N_s$ samples is $O(N_sN^3/3)$. For the BAS, assuming that there are $N_{u,i}$ unique samples generated at the $i$-th step (which is the width of the $i$-th layer in Fig.~\ref{fig:bas}(b)), then the total computational cost is $\sum_{n=1}^N i^2 N_{u,i}$. If we assume that $N_{u,i}$ is roughly a constant, denoted as $\uniquesample$ ($\uniquesample$ is usually less than $10^6$ even for molecular systems with $N>100$), then the cost of BAS is only $O(\uniquesample N^3/3)$, which is independent of $N_s$. 
Existing researches on the acceleration of training~\cite{lightseq2-sc22} and inference~\cite{lightseq-infer,turbotransformers} for Transformer models on GPU platforms can be leveraged to design the GPU-accelerated implementation for the sampling algorithm.

\subsection{Data-centric parallelization strategy for NNQS} \label{sec:para}

\begin{figure}[!htbp]
	\centerline{\includegraphics[width=\columnwidth]{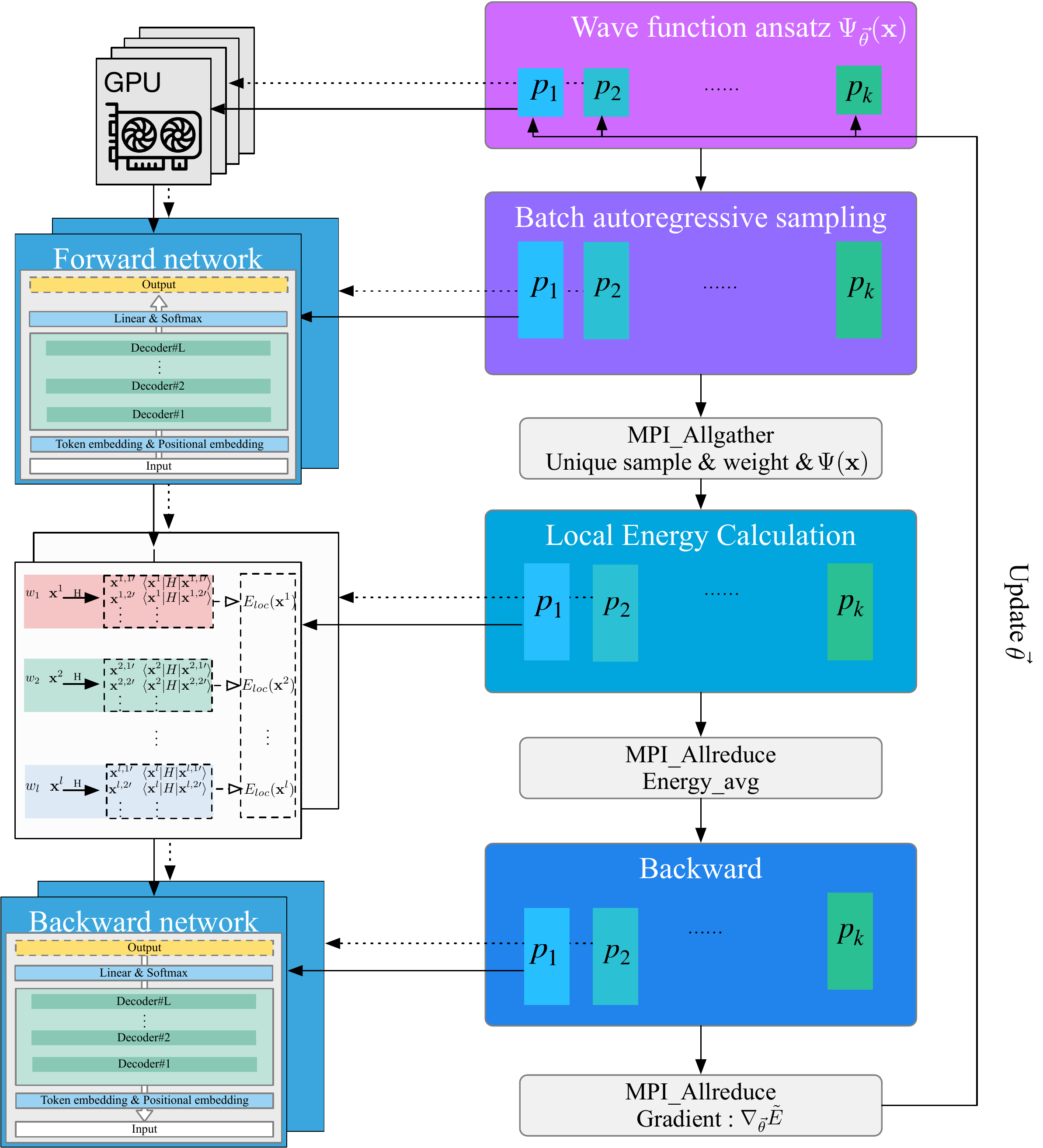}}
	\caption{Schematic of the data-centric parallelization scheme for NNQS using autoregressive wave function ansatz. Each process in the first level parallelization takes charge of a number of unique samples with their weights throughout the computation.}
	\label{fig:para_flow}
\end{figure}

We propose a customized data-centric parallelization scheme for the VMC algorithm which is valid for all autoregressive neural networks, as demonstrated in Fig.~\ref{fig:para_flow} and in comparison with the serial scheme in Fig.~\ref{fig:serial_flow}. 
Our parallelization scheme contains two levels.
In the first level parallelization, we divide all the unique samples~($\uniquesample$) into $N_p$ batches and 
parallelize the sampling, local energy and gradient calculation (backward propagation) over those batches, such that each functions processes
$\uniquesample/N_p$ unique samples (instead of $N_s/N_p$ samples).
In the second level parallelization, the sampling and backward propagation functions for each batch are further parallelized using pytorch multi-threading, while for local energy we use our parallel local energy evaluation scheme (see Sec.~\ref{sec:localenergy}) over $N_t$ threads by further divide the $\uniquesample/N_p$ samples into $N_t$ smaller batches.

More concretely, our parallelization scheme contains $6$ stages in each VMC iteration: 1) We use a customized parallel BSA strategy (see Sec.\ref{sec:pbsa}) over a total of $N_p$ processes to obtain approximately $\uniquesample/N_p$ unique samples in each process; 2) We use MPI\_Allgather to synchronize all the samples on each process, with $\uniquesample N_p(\lceil N/8\rceil+16)$ Byte data communication; 3) Each process takes a chunk of $\uniquesample/N_p$ unique samples and performs the parallel local energy evaluation scheme; 4) Collecting all the local energies together into the main process, computing the average therein and then broadcasting it to all the processes using MPI\_Allreduce, with $16N_p$ Byte data communication; 5) Performing backward propagation on each process independently; 6) Collecting all the local gradients from each process into the main process and calculating the weighted average, then updating the parameters using the gradient-based optimizer and send the new parameters to all the processes using MPI\_Allreduce, with $8M N_p$ Byte for $M$  parameters data communication.
Importantly, during the whole process of our parallel VMC iteration, only steps 2), 4), 6) involve very minimal data communication (Typically C$_2$ with STO-3G basis set, $N=20$, $N_u=2.7\times10^4$, $N_p=64$ and $M=2.7\times 10^5$, the total data communication volume is about $173$ MB in our one iteration calculations), and the unique samples are always kept in their processes until all calculations finish (therefore data-centric).

\subsection{Parallel batch autoregressive sampling algorithm}\label{sec:pbsa}

\begin{figure}[!htbp]
\centerline{\includegraphics[width=\columnwidth]{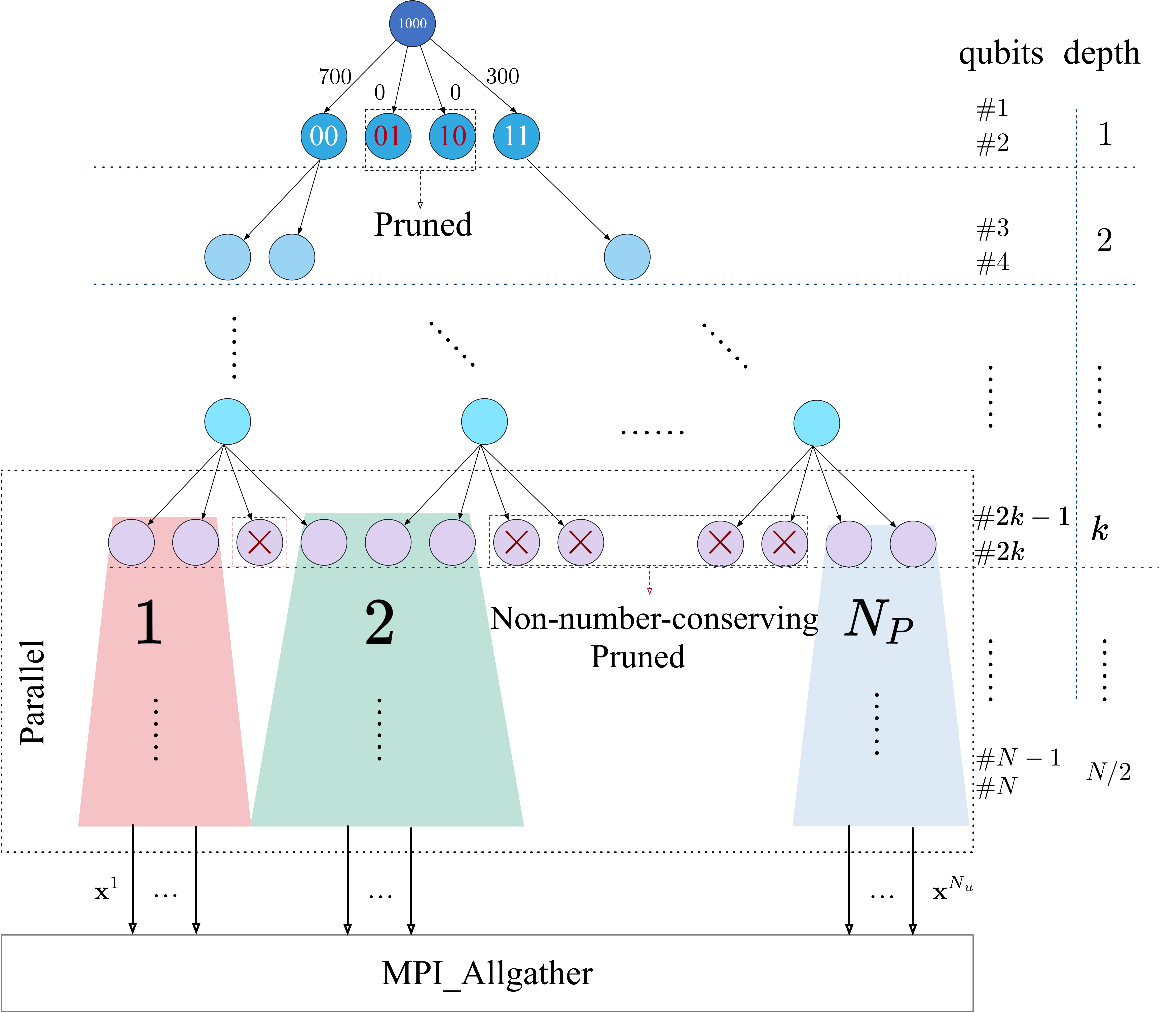}}
\caption{The parallel batch autoregressive sampling algorithm. 
 The tasks are divided at the $k$-th local sampling step ($k$-th layer of the tree) such that the number of samples on each process are approximately equal. Pruning is done in each step to remove leaves with zero occurrence and which break the number conservation constrain.}

\label{fig:sampling}
\end{figure}

The first challenge in NNQS is the sampling process. 
Autoregressive neural networks allow one to generate exact samples without the need of a pre-thermalization step and throwing away intermediate samples (to avoid auto-correlations) necessary for MC sampling, thus are more efficient than MC sampling in general. Moreover, the batch sampling algorithm can output a large number of unique samples together with their occurrences in a single run (see Fig.~\ref{fig:bas}). 
However paralleling the BSA on distributed architectures is not straightforward, mainly because 1) in BSA the calculation in one local sampling step is dependent on all the previous steps and 2) the positions and the total number of the unique samples in each step are not known before hand. 

\begin{figure*}
\centerline{\includegraphics[width=2\columnwidth]{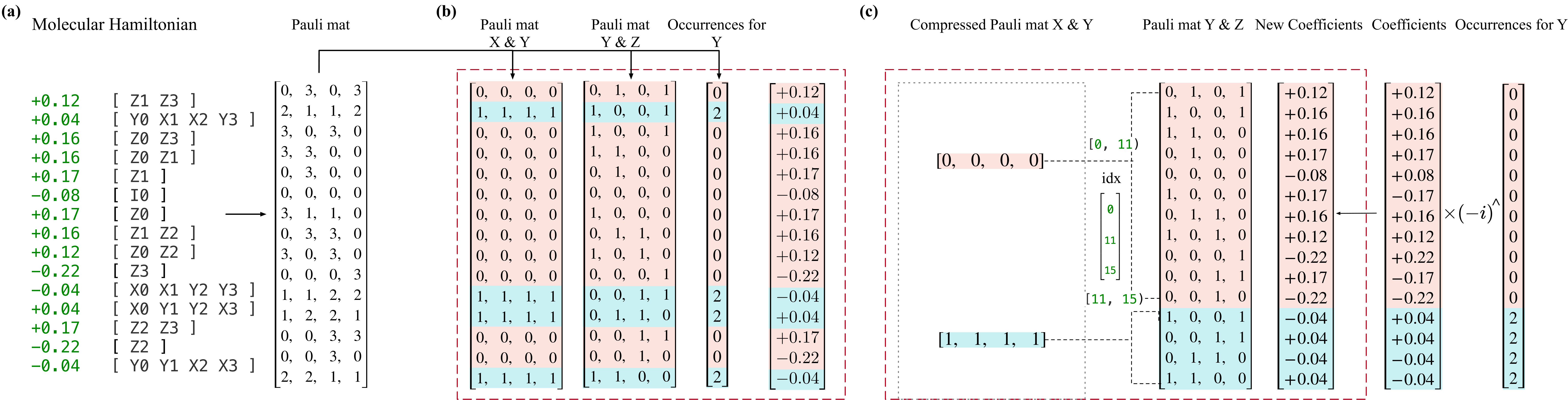}}
\caption{(a) Hamiltonian for the H$_2$ molecular system with $4$ qubits and $2$ electrons stored as a list of Pauli strings. 
(b) The scheme used in Ref.~\cite{Zhao_2023} to store the Hamiltonian in memory. (c) Our compressed data structure to store the complete information of the Hamiltonian. The data inside the dashed red boxes are the information stored in memory in both (b) and (c).}
\label{fig:ham}
\end{figure*}

To solve this issue,
we propose a heuristic parallelization strategy for BSA that proves to be effective in practice, which is shown in Fig.~\ref{fig:sampling}. We represent the whole BSA process as a tree where each layer corresponds to a local sampling step.
We sample two qubits each time (thus a quadtree) since they correspond to the same spatial orbital. We also sample in the reverse order of the qubits after the JW transformation as in Ref.~\cite{Zhao_2023}. We first perform the serial BSA on each process using the same random seed (such that we get exactly the same samples on each process) for the first $k$ steps, and then we divide the unique samples (nodes in Fig.~\ref{fig:sampling}) at the $k$-th step into $N_p$ processes such that each process gets approximately the same number of samples (instead of unique samples) in this step. This does not guarantee that each process generates the same number of unique samples in the end (thus achieving good load balance since the later calculations are directly determined by the number of unique samples), but from our numerical experiments we observe that approximate load balance can be achieved with this strategy if $k$ is chosen properly. In our implementation we use a simple strategy to dynamically determine $k$: we set a threshold $\uniquesample^{\ast}$ and choose $k$ to be the first local sampling step such that the current number of unique samples~( $N_{u,k}$) is larger than $\uniquesample^{\ast}$.
We note that in this strategy the calculations from the $1$ to $k-1$-th steps are simply repeated on each process, however, this will not significantly affect the overall computational cost since the computational cost of the local sampling grows as $i^2$ against step $i$.

During each step, we prune the leaves with $0$ occurrence. In addition, there is a number conserving constrain which can be used to reduce the sample space size: the total number of spin up and spin down electrons, denoted as $n_{\uparrow}$ and $n_{\downarrow}$, are conserved separately. During the JW transformation, the $i$-th spatial orbital is mapped to the two qubits at positions $2i-1$ and $2i$, corresponding to the two spin orbitals with up and down spins. Denoting each $x_i = y_{2i}y_{2i-1}$ with $y_j \in \{0, 1\}$ (since we sample two qubits in each step), this number conservation constrain can be imposed by regularizing the local conditional probability distributions $\pi(x_i|x_{i-1}, \dots, x_1)$ as~\cite{Zhao_2023}:
\begin{align}
\tilde{\pi}(x_i|x_{i-1}, \dots, x_1) = \begin{cases} 
0, \text{if } \sum_{j=1}^{i} y_{2j-1} > n_{\uparrow} \\
0, \text{if } \sum_{j=1}^{i} y_{2j} > n_{\downarrow} \\
\pi(x_i|x_{i-1}, \dots, x_1), \text{otherwise}
\end{cases},
\end{align}
and then using $\tilde{\pi}(x_i|x_{i-1}, \dots, x_1) / \sum_{x_i} \tilde{\pi}(x_i|x_{i-1}, \dots, x_1)$ instead.

\subsection{Accelerating local energy calculation on GPU}\label{sec:localenergy}

The second challenge in NNQS for electronic structure calculations is to evaluate the local energy in Eq.(\ref{eq:Eloc}).  For molecular systems this step could be very expensive due to the $O(N^4)$ scaling of the total number of Pauli strings. 
This could also lead to a memory issue since there are $O(N^4)$ possible $\boldx'$ coupled to an input $\boldx$.
We have used several techniques to parallelize the computation on a GPU while at the same time limit the memory usage and reduce the computational cost, such that we can scale up our calculations to more than $100$ qubits. These techniques are shown as follows.

\begin{figure}[!htbp]
\centerline{\includegraphics[width=\columnwidth]{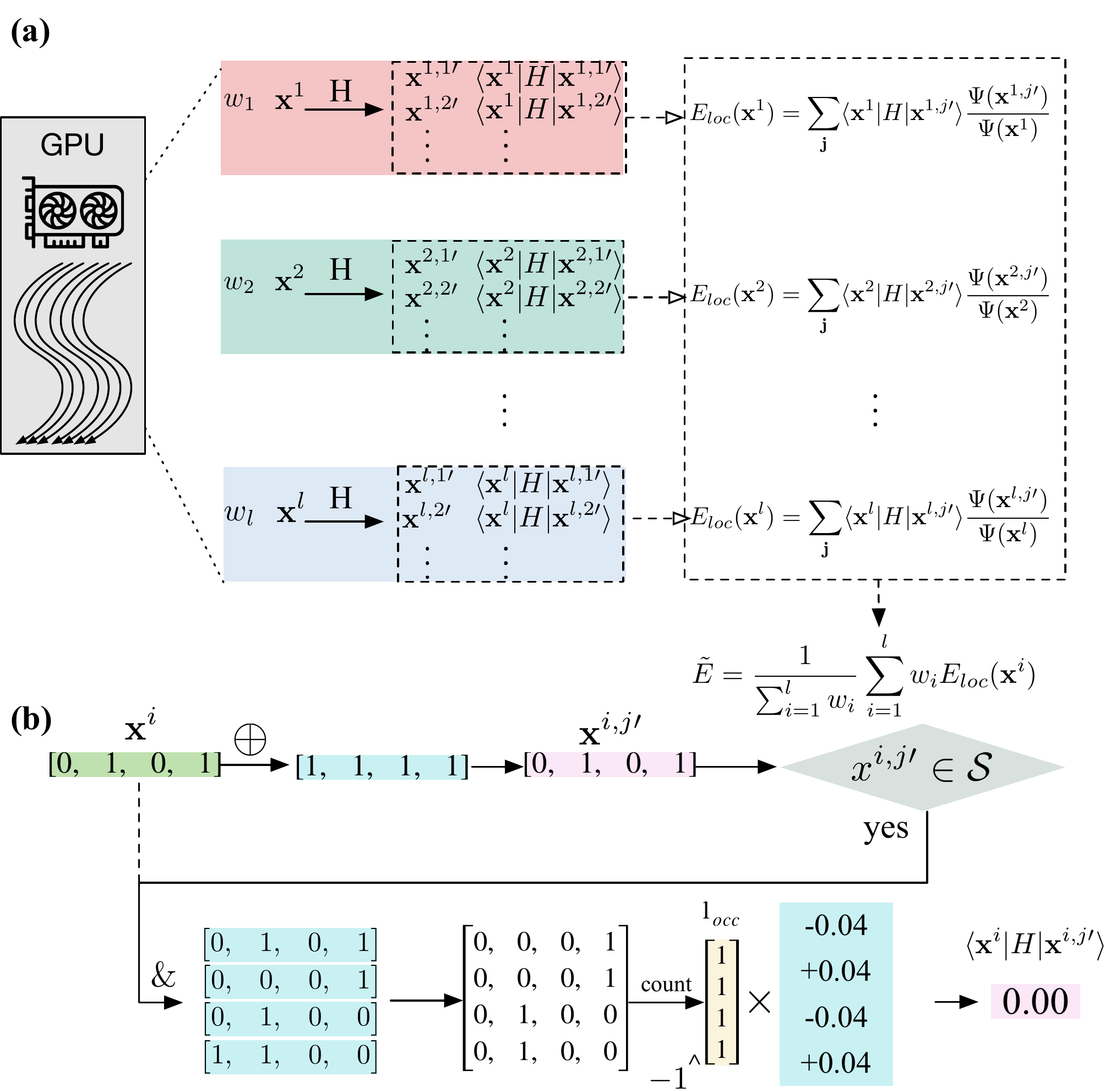}}
\caption{(a) Parallelizing the calculation of local energies over the samples. (b) Calculating the local energy for a given input sample.}
\label{fig:loc_energy}
\end{figure}

(1) A highly compressed data structure to store the Hamiltonian in memory. For large molecular systems storing the Hamiltonian as a list of symbolic Pauli strings (or integers, see Fig.~\ref{fig:ham}(a)) would take a large amount of memory and be inefficient for later calculations. In Ref.~\cite{Zhao_2023} a compressed data structure is used to represent the Hamiltonian, where each Pauli string is represented by three components: 
two boolean tuples of length $N$ where the first one stores the occurrence of $X$ or $Y$ and is used to calculate the states $\boldx'$ coupled to each $\boldx$, while the second one stores the occurrence of $Y$ or $Z$ and is used to calculate the coefficient, 
plus an integer to store the occurrence of $Y$ used to calculate the sign. This data structure is shown in Fig.~\ref{fig:ham}(b). 
We propose a more memory-efficient scheme as shown in Fig.~\ref{fig:ham}(c), where we only store the unique ones of the first list of boolean tuples and reorganize the second list accordingly. The coefficient is precomputed using the occurrence of $Y$ and updated in-place. The computational cost of the reorganization step can be neglected since it only needs to be done once and can be pre-computed. The memory reduction of our scheme in Fig.~\ref{fig:ham}(c) compared to the scheme in Fig.~\ref{fig:ham}(b) is around $40\%$ in general (see Fig.~\ref{fig:ham-mem-opt} in Sec.~\ref{sec:speedup}). In addition, for each input $\boldx$, all the Pauli strings with the same first boolean tuple will couple $\boldx$ to the same bitstring $\boldx'$, therefore our data structure naturally allows one to calculate each unique $\boldx'$ only once (the coefficients for duplicate $\boldx'$ are simply summed over as shown in Fig.~\ref{fig:loc_energy}(b)).
Algorithm~\ref{alg:preprocess} meticulously delineates the aforementioned optimizations, the initial step entails the parsing of inputted Pauli strings, which serves to construct Pauli matrices XY and YZ. Simultaneously, occurrences of Y are enumerated to compute the new coefficients. Following this, a dictionary data structure (Dict) and state encoding are employed to eliminate redundant states. Lastly, continuous memory space is allocated. The dictionary-type data is compressed into this continuous space for storage, and corresponding ranges are retained using an index array (idxs).

(2) A fused nonzero Hamiltonian entry evaluation and local energy calculation design. That is, when evaluating $\Eloc(\boldx)$ using Eq.(\ref{eq:Eloc}), 
instead of finding all the different $\boldx'$ with nonzero $H_{\boldx \boldx'}$ (together with their coefficients) and then using them to evaluate $\Eloc(\boldx)$, 
we directly calculate the contribution of each unique $\boldx'$ to $\Eloc(\boldx)$ (namely $H_{\boldx \boldx'}\Psi_{\para}(\boldx')/\Psi_{\para}(\boldx)$ which is a scalar) as soon as it is found, and then sum over them. In this approach we can calculate $\Eloc(\boldx)$ by only storing two bitstrings in memory.

\begin{algorithm}
\LinesNumbered
\KwIn{pauliStrings, n\_qubits}
\KwResult{idxs, pmXY\_buf, pmYZ\_buf, coeffs\_buf}

{$N, K \gets \text{n\_qubits}, \text{length(pauliStrings)}$\\}
$\text{pmXY, pmYZ} \gets \text{zeros}(N,K), \text{zeros}(N,K)$\\

\For{($i$, \rm{(coeff, pauliStr)}) \textbf{in} \rm{enumerate(pauliStrings)}}{
    $Y_{occ} \gets 0$ \tcp{occurrence for Y}
    \tcc{Construct Pauli mat XY and Pauli mat YZ}
    \For{($p$, \rm{pauli}) \textbf{in} \rm{pauliStr}}{
        \If{\rm{pauli} $ == "X"$}{
            pmXY[$p$, $i$] $\gets 1$\\
        }
        \ElseIf{\rm{pauli} $ == "Y"$}{
            pmXY[$p$, $i$] $\gets$ pmYZ[$p$, $i$] $\gets 1$\\
            $Y_{occ} \gets Y_{occ} + 1$\\
        }
        \Else{pmYZ[$p$, $i$] $\gets 1$}
    }
    \tcc{New coefficients: fuse const calculation}
    coeff $\gets \text{real}(coeff) \times \text{real}((-i)^{Y_{occ}})$ \\

    \tcc{Compressed Pauli mat XY using Dict}
    sid = \_encode2id(pmXY[:,$i$])\\
    \If{\textbf{not} \rm{haskey(coeffs\_dict, sid)}}{
        pmXY\_dict[sid] $\gets$ pmXY[:,$i$]\\
    }
    coeffs\_dict[sid].append(coeff)\\
    pmYZ\_dict[sid].append([pmYZ[:,$i$]])\\
}

\tcc{Compacting data into new continuous buffer}

\For{($i$, \rm{(sid, pm23)}) \textbf{in} \rm{enumerate(pmYZ\_dict)}}{
    $l\gets$ \text{length}(pm23)\\
    pmYZ\_buf[:,$j:j+l$] $\gets$ pm23 \\
    pmXY\_buf[:, $i$] $\gets$ pmXY\_dict[sid]\\
    coeffs\_buf[$j$:$j$+$l$] $\gets$ coeffs\_dict[sid] \\
    $j \gets j+l$; idxs[$i$] $\gets j$\\
}
\caption{Preprocess: Hamiltonian data structure optimizations}
\label{alg:preprocess}
\end{algorithm}

(3) Parallelizing the calculation of local energies over unique samples on a GPU. 
Each GPU is responsible for all the calculations related to a batch of $l$ unique samples ($l = \uniquesample/N_p$).
GPU parallelization is implemented by dividing the $l$ unique samples into $N_t$ smaller batches (each containing $l/N_t$ unique samples) and calculating the local energies of each batch on one thread, which is shown in Fig.~\ref{fig:loc_energy}.

(4) A sample-aware scheme for evaluating $\Eloc(\boldx)$. During the local energy calculation, all the unique samples, denoted as $\samples = \{\boldx^1, \boldx^2, \cdots, \boldx^{\uniquesample}\}$, are stored on each GPU (see Sec.~\ref{sec:para}). When evaluating $\Eloc(\boldx)$, we calculate the coefficient $H_{\boldx \boldx'}$ only if $\boldx' \in \samples$, as illustrated in Fig~\ref{fig:loc_energy}(b) (see Fig.~\ref{fig:speedup-eloc} in Sec.~\ref{sec:speedup} for actual speedup using this technique). 

(5) Efficient storage of the unique samples $\samples$ as a lookup table. The $l$ samples generated on each GPU are stored as boolean lists for computational efficiency as can be seen from Fig.~\ref{fig:loc_energy}(b). However, when synchronizing the unique samples on all the GPUs, a memory efficient storage scheme is preferable to reduce data communication, therefore we simply encode the boolean tuple into the bits of a $64$-bit integer (in case $64\leq N<128$, we use two integers). This storage format also allows us to use the very efficient binary search during the sample-aware local energy evaluation, since the integers can be naturally ordered according to their values.

Algorithm~\ref{algo:eloc-optim} includes optimization techniques (2)-(5), each of which assigns a subset of samples to a single GPU thread. The computation process prioritizes coupled states. Within the identification lookup table (id\_lut), a binary search is conducted. If a matching state exists, the computation proceeds to calculate the coupling coefficients. By accessing the wavefunction lookup table (wf\_lut), the necessity for forward inference in the neural network is bypassed, allowing for immediate contributions to be aggregated. In practical implementation, complex numbers are separated into their real and imaginary components for independent computation, and the resultant values are subsequently written into their respective positions.

\begin{algorithm}
\DontPrintSemicolon
\LinesNumbered
\SetAlgoLined
\KwIn{n\_qubits, $NK$, idxs, coeffs, pmXY, pmYZ, batch\_size\_cur\_rank, $ist$, states, id\_lut, wf\_lut}
\KwResult{res\_elocs}
\tcc{Initialization of power of two table}
tbl\_pow2$[0...63] \leftarrow \{1, 2,..., 2^{63}\}$ \;
\tcc{Partition for CUDA kernel}
index $\leftarrow$ blockIdx.x $\times$ blockDim.x $+$ threadIdx.x \;
stride $\leftarrow$ gridDim.x $\times$ blockDim.x \;

\For{$ii \leftarrow$ \rm{index} \KwTo \rm{batch\_size\_cur\_rank} \textbf{step} \rm{stride}}{
    eloc\_real, eloc\_imag $\leftarrow 0$ \;
    \For{$k \leftarrow 0$ \KwTo $NK$}{
        \tcc{Fuse coupled states and state2id}
        id $\leftarrow 0$ \;
        \For{$j \leftarrow 0$ \KwTo $N$}{
            id $+=$ xor(states[$ii \times N+j$], pmXY[$k \times N+j$]) $\times$ tbl\_pow2[$j]$ \;
        }
        \tcc{Binary find id among the lookup table}
        idx $\leftarrow$ binary\_find(id\_lut, id) \;
        \If{\rm{idx} \textbf{does not exist}}{
            \textbf{continue} \;
        }
        \tcc{Calculate coupled coef}
        coef $\leftarrow 0.0$ \;
        \For{$i \leftarrow \rm{idxs}[k]$ \KwTo $\rm{idxs}[k+1]$}{
            $\rm{sum} \leftarrow 0$ \;
            \For{$j \leftarrow 0$ \KwTo $N$}{
                sum $+=$ states[$ii \times N+j$] \& pmYZ[$i \times N+j]$ \;
            }
            coef $+=$ ( (sum \& $1) ? 1 : -1) \times$ coeffs[$i]$ \;
        }
        \tcc{Accumulation}
        eloc\_real $+=$ coef $\times$ wf\_lut[idx$\times 2]$ \;
        eloc\_imag $+=$ coef $\times$ wf\_lut[idx$\times 2+1]$ \;
    }
    \tcc{Store the result number as return}
    a, b $\leftarrow$ eloc\_real, eloc\_imag \;
    c, d $\leftarrow$ wf\_lut[$(ist+ii) \times 2$], wf\_lut[$(ist+ii) \times 2+1$] \;
    c2\_d2 $\leftarrow$ c$^2$ + d$^2$ \;
    res\_elocs[$ii \times 2] \leftarrow$ (a $\times$ c + b $\times$ d) $/$ c2\_d2 \;
    res\_elocs[$ii \times 2+1] \leftarrow - $(a $\times$ d $-$ b $\times$ c) $/$ c2\_d2 \;
}
\caption{Optimizations of local energy calculation}
\label{algo:eloc-optim}
\end{algorithm}

\section{Evaluation}

\subsection{Experimental setup}
The wave function ansatz used in all the simulations in this work is set as follows. For the amplitude part, we have used two decoders with $d_{model}=16$ (embedding size) and $n_{heads}=4$ (number of heads). For the phase part, we have used three dense layers in the MLP with sizes $N\times 512 \times 512\times1$. 
We achieved optimal outcomes exploring MLP layers (3-5), hidden layers (256-1024), and decoder layers (2-4) with a $d_{model}$ range of 16-64.
We have used the gradient descent optimizer AdamW for training with the learn rate schedule 
\begin{align}
\alpha_i=d_{model}^{-0.5} \times \min(i^{-0.5}, i\times S_{warmup}^{-1.5}),
\end{align}
where $\alpha_i$ is the learn rate of the $i$-th training epoch and we set the warm up steps as $S_{warmup}=4000$. The maximum number of VMC iterations is set to $10^5$. 
All our simulations are performed on AMD EPYC 7742 CPU and NVIDIA A100 PCIe 80GB for CPU and GPU computation environment respectively.


During the training process, we start from a wave function ansatz with randomly initialized parameters.
In the pre-training stage, we observe that a large number of unique samples will be generated since the wavefunction is far from exact, thus we set a lower threshold $N_s$ ($N_s=10^5$ in the first $100$ VMC iterations) for the number of samples to increase efficiency. In the later stage we set the threshold (gradually increase $N_s$ until it reaches the maximum value of $10^{12}$) to be very large for accurate calculation (we observe that in the later stage, only a smaller number of unique samples will be generated). The efficacy of the model is assessed based on convergence precision.

\subsection{Simulation Validation }	  	

\begin{table*}[h!]
\centering
\begin{tabular}{ccccccccc|c}
\hline
\hline
Molecule & $N$ & $N_e$ & $N_h$ & HF  & CCSD & NAQS~\cite{BarrettLvovsky2022} & MADE~\cite{Zhao_2023} & \wyj{QiankunNet}  & FCI \\ \hline
H$_2$O  & 14 & 10 & 1390 & $-74.964$  & -75.0154   & -75.0155 & -75.0155  & -75.0155 & -75.0155   \\ \hline
N$_2$  & 20 & 14  & 2239 & $-107.4990$ & -107.6560 & -107.6595 & -107.6567  & -107.6601 & -107.6602   \\ \hline
O$_2$    & 20 & 16  & 2879 & $-147.6319$ & -147.7027  & -147.7500 & -147.7499 & -147.7501 & -147.7502   \\ \hline
H$_2$S  & 22 & 18 & 9558 & $-394.3114$ & -394.3545 & -394.3546 & -394.3545 & -394.3546  & -394.3546   \\ \hline
PH$_3$  & 24 & 18  & 24369 & $-338.6341$ & -338.6981 & -338.6984 & -338.6981 & -338.6982 & -338.6984   \\ \hline
LiCl  & 28 & 20 & 24255 & $-460.8273$ & -460.8475  & -460.8496 & -460.8481 & -460.8494 & -460.8496   \\ \hline
Li$_2$O  & 30 & 14 & 20558 & $-87.7956$ & -87.8855  & -87.8909 & -87.8856 & -87.8907 & -87.8927   \\ \hline 
MAE (Hartree)  &  &  &  &  &  & $3.8\times 10^{-4}$ & $1.8\times 10^{-3}$ & $3.7\times 10^{-4}$ &    \\ \hline \hline
\end{tabular}
\caption{Ground state energies (in Hartree) calculated by our method. The conventional HF and CCSD results, FCI results as well as existing NNQS results including NAQS~\cite{BarrettLvovsky2022} and MADE~\cite{Zhao_2023} are shown for comparison. $N$ is the number of qubits, $N_e$ the total number of electrons (including spin up and spin down) and $N_h$ the total number of Pauli strings of the Hamiltonian. The mean absolute errors (MAE) for various methods compared with FCI are also listed. }
\label{tab:validation}
\end{table*}

To demonstrate the precision of our transformer-based NNQS method, we first compute the ground state energies of several small-scale molecular systems and compare them to existing results using NNQS, which are shown in Table.~\ref{tab:validation}.
The mean absolute error (MAE) are 
listed for each method. 
We can see that our method can reach about the same precision as NAQS (for N$_2$ we are more precise) and is generally more accurate than MADE.

\begin{figure}[!htbp]
\centerline{\includegraphics[width=\columnwidth]{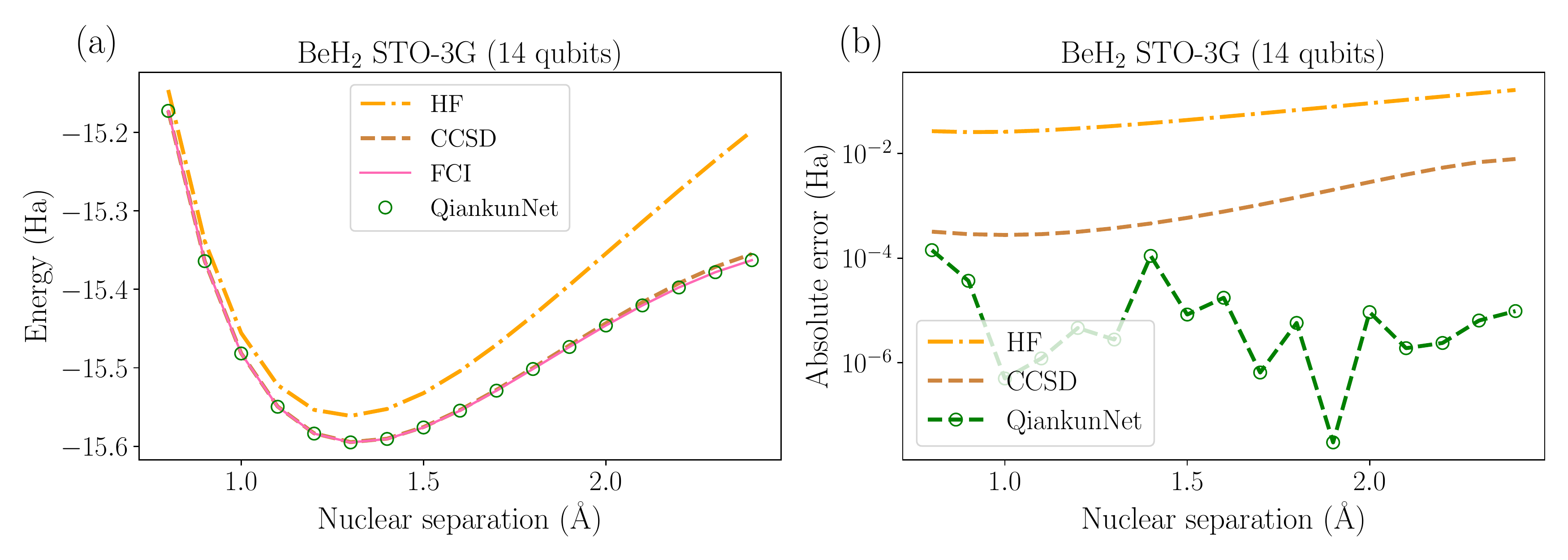}}
\caption{(a) Potential energy surface of the BeH$_2$ molecular system in the STO-3G basis set calculated by our method (green circles), compared to the HF, CCSD and FCI energies. (b) The absolute errors with respect to FCI results.}
\label{fig:pes-beh2}
\end{figure}

We also apply our method to calculate the potential energy surface of the BeH$_2$ molecular system in the STO-3G basis set ($14$ qubits),
which is shown in Fig.~\ref{fig:pes-beh2}. We can see that chemical accuracy (within an absolute error of $1.6\times 10^{-3}$ Hartree compared to FCI) can be reached in the potential energy surface simulation.

\subsection{Memory saving and speedup of local energy calculation}\label{sec:speedup}

\begin{figure}[!htbp]
\centerline{\includegraphics[width=\columnwidth]{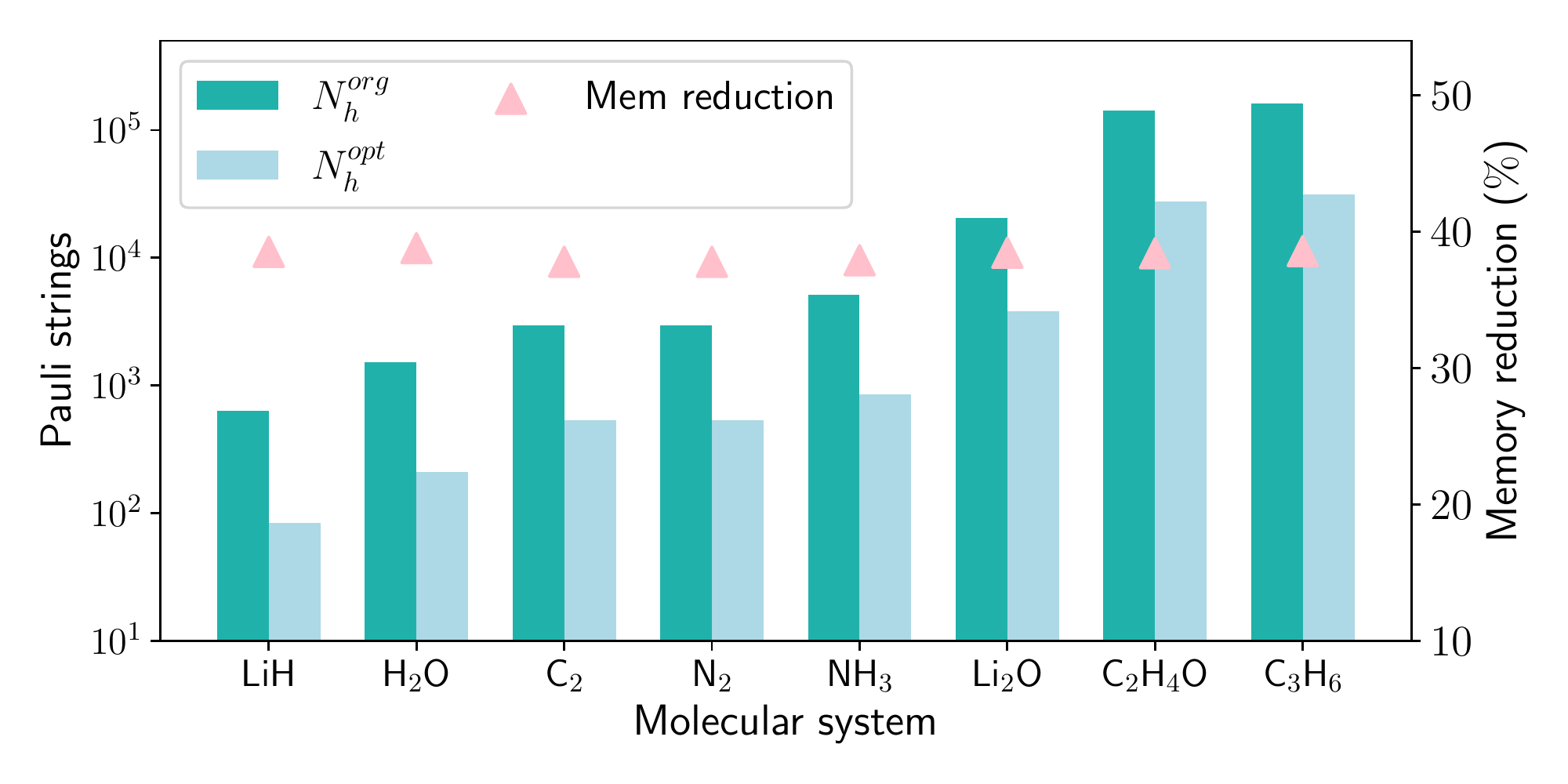}}
\caption{Memory reduction using our compressed data structure to store the Hamiltonian compared to the original scheme from Ref.~\cite{Zhao_2023} for several exemplary molecular systems in the STO-3G basis set.}
\label{fig:ham-mem-opt}
\end{figure}

Now we show the memory reduction of using our compressed data structure for the Hamiltonian (see Fig.~\ref{fig:ham}(c) and Sec.~\ref{sec:localenergy}), compared to the scheme proposed in Ref.~\cite{Zhao_2023} (see Fig.~\ref{fig:ham}(b)), which is plotted in Fig.~\ref{fig:ham-mem-opt} for a wide variety of molecular systems. We can see that generally there is a more than $40\%$ memory reduction.

\begin{figure}[!htbp]
\centerline{\includegraphics[width=\columnwidth]{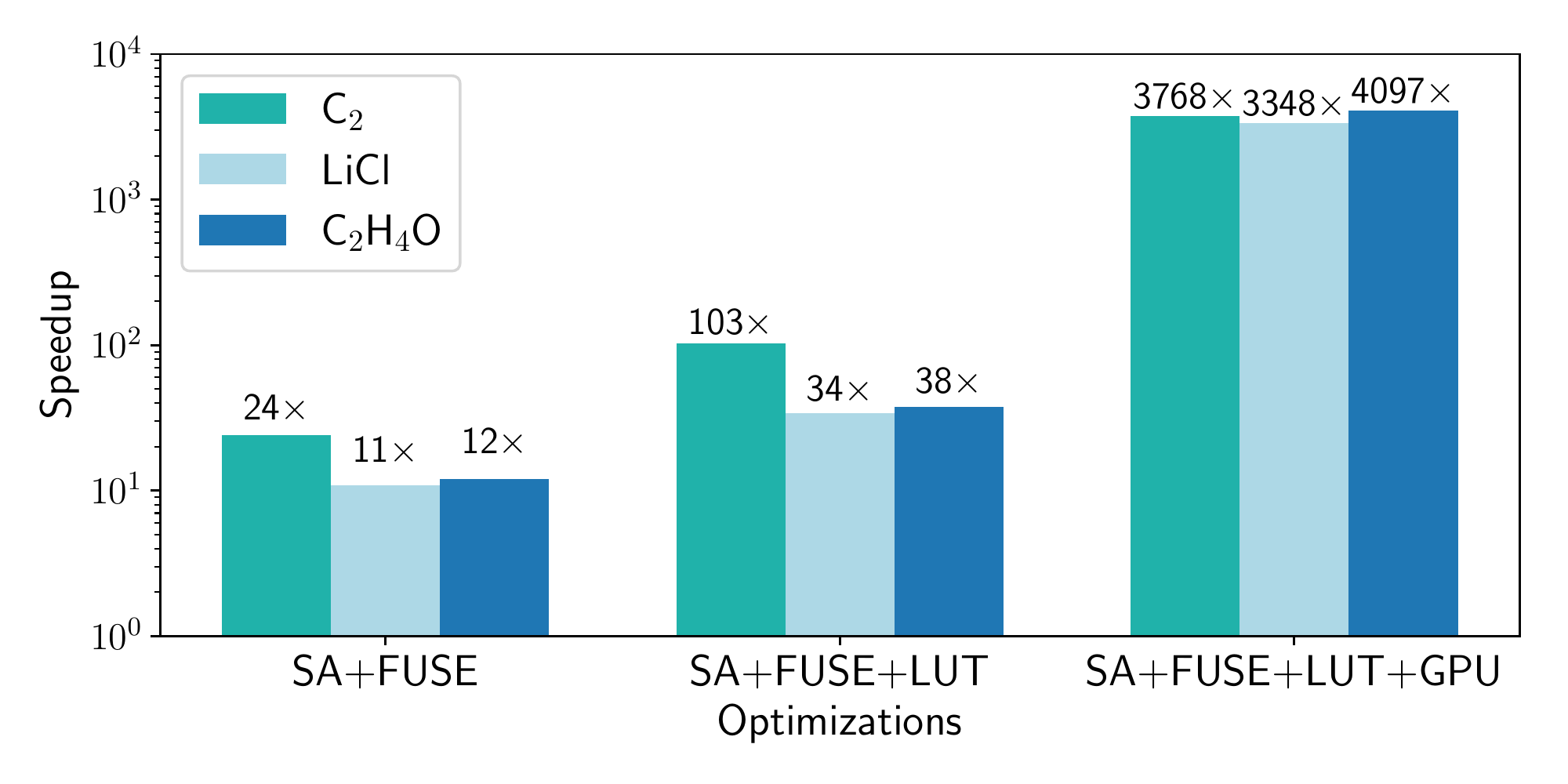}}
\caption{The runtimes for local energy calculation by integrating different optimization techniques, for the C$_2$, LiCl and C$_2$H$_4$O molecular systems in the STO-3G basis set with $N=20,28,38$. The total number of unique samples produced in these simulations are $\uniquesample = 10553, 9271, 25622$ respectively. SA means the sample-aware evaluation scheme (Sec.~\ref{sec:localenergy}, method (4)), FUSE means the fused design of nonzero Hamiltonian entry evaluation and local energy calculation (Sec.~\ref{sec:localenergy}, method (2)), LUT means to store the unique samples as an efficient lookup table (Sec.~\ref{sec:localenergy}, method (5)), GPU means the local energy calculation implemented on GPU (Sec.~\ref{sec:localenergy}, method (2)). 
The baseline is a bare CPU version without using the optimization techniques mentioned here.
The other techniques in Sec.~\ref{sec:localenergy} which are not mentioned here are used in all tests. 
}
\label{fig:speedup-eloc}
\end{figure}

In Fig.~\ref{fig:speedup-eloc}, we show the speedups of the local energy calculation induced by different optimization techniques, for the C$_2$, LiCl, C$_2$H$_4$O molecular systems. Taking C$_2$ as an example,
first, the sample aware evaluation scheme (reducing the computational cost) plus the fused design of nonzero Hamiltonian entry evaluation and local energy calculation (increasing data locality and reducing memory usage), yields a performance gain of $24$x. Second, by storing the unique samples as an ordered integer list, which reduces memory usage and increases search efficiency, the performance gain further increases to $103$x. Third, by using GPU parallelization, the overall performance gain becomes $3768$x compared to the baseline CPU version where none of these optimizations are used.
Overall, implementing our parallel local energy calculation scheme on GPU gives us around $4,000$x speed up compared to the baseline CPU version, which is essential to scale up the simulation scale.

\subsection{Scalability Results}\label{subsec:parallel performance}

\begin{figure}[!htbp]
\centerline{\includegraphics[width=\columnwidth]{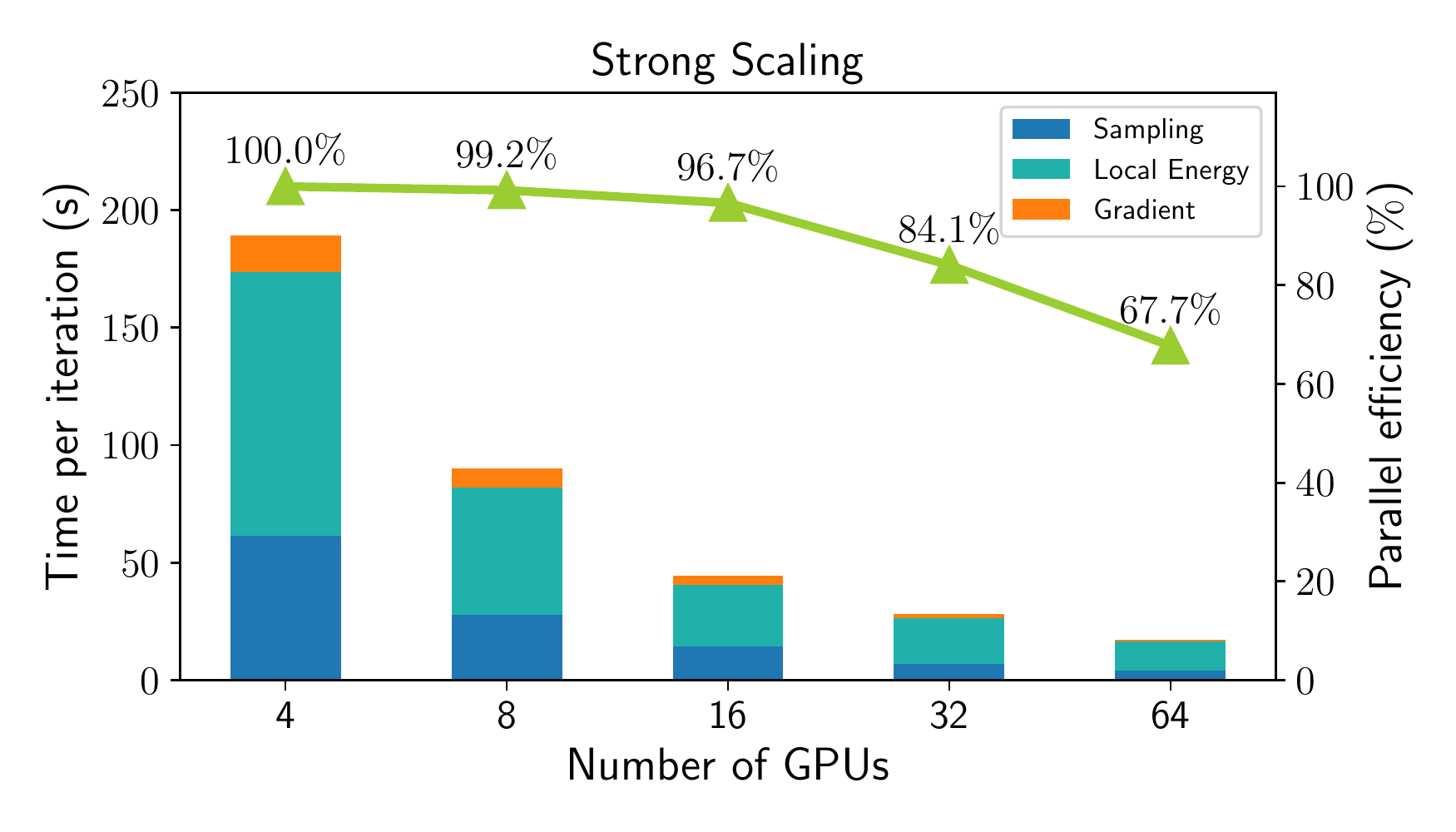}}
\caption{Strong scaling of the computation for the benzene molecular system in the 6-31G basis set ($120$ qubits). The green line shows the strong parallelization efficiency. Three functions (sampling, local energy calculation, backpropagation) are profiled in the computation time.}
\label{fig:strong}
\end{figure}

We use the benzene molecular system in the 6-31G basis set with $120$ qubits for the scalability test, for which $N_h = 2434919$. We measure the performance after the pre-training stage such that the number of unique samples in each iteration becomes more or less stable. We have used $\uniquesample^{\ast} = 16384 n$ for $n$ GPUs in the parallel BAS for both the strong and weak scaling test.

Fig.~\ref{fig:strong} shows the strong scaling performance of our method, where we have used $N_s=1.6\times 10^6$ and obtain $\uniquesample\approx 6.5\times 10^5$. 
As the number of used GPUs increases from $4$ to $32$, the strong parallelization efficiency is still higher than $84\%$. While when increased to $64$ GPUs, the efficiency decreases to $67.7\%$.
This slightly lower parallel
efficiencies can be attributed to two factors: First the relatively small number
of unique samples ($\approx 10^4$) that is allocated on each GPU when distributing $6.5\times 10^5$ unique samples over $64$ GPUs, the parallelization efficiency of GPU is not well utilized at this scale; Second, during the  parallel BAS process, the imbalance of the number of unique samples generated on each GPU could happen due to the pruning of the samples with physical constrains.
We note that the later the parallel BAS is performed, the more evenly distributed the unique samples will be, yet the time for prior serial sampling process will increase.

\begin{figure}[!htbp]
\centerline{\includegraphics[width=\columnwidth]{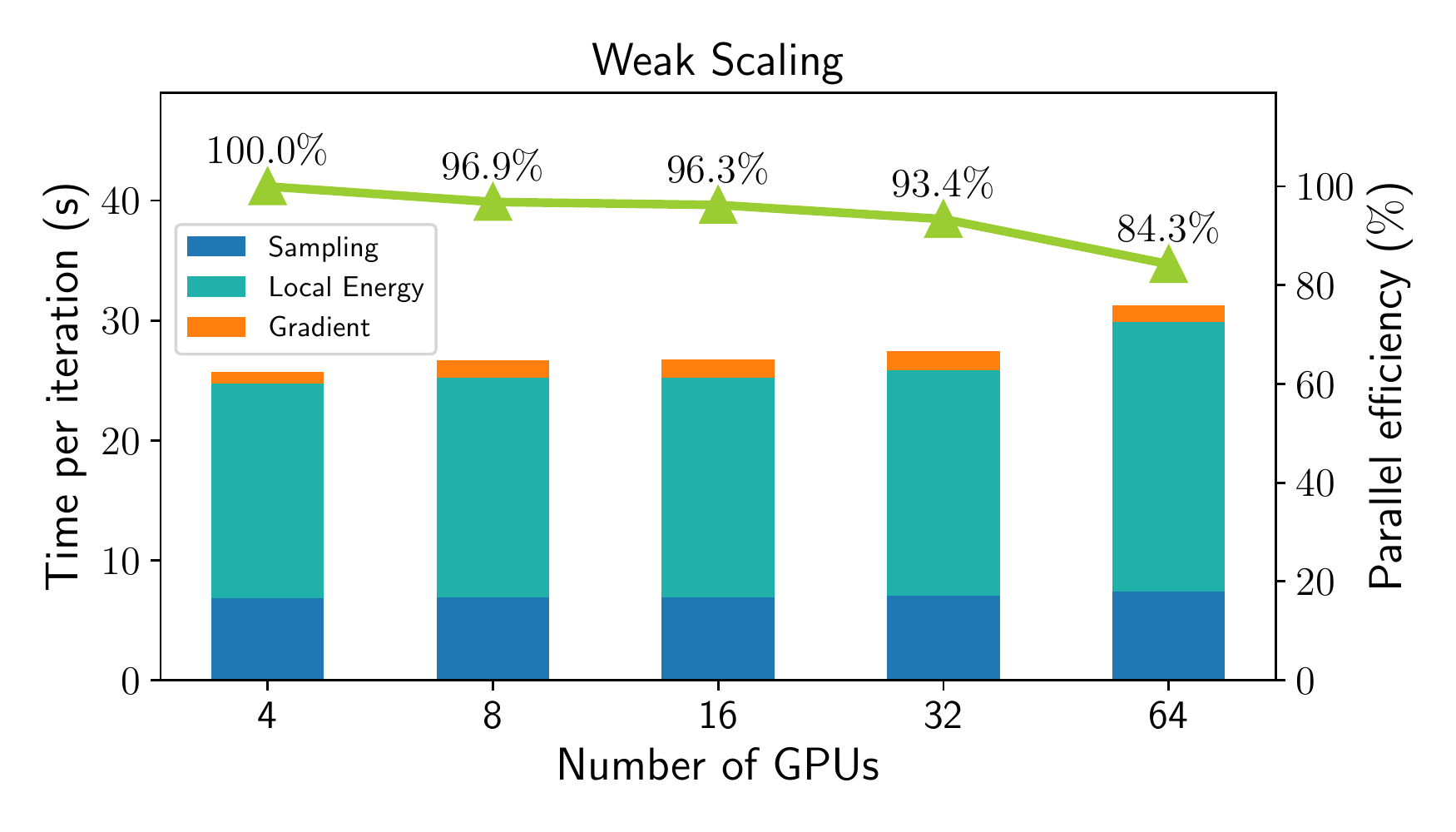}}
\caption{Weak scaling of the runtime for the benzene molecular system in the 6-31G basis set. To ensure that a approximately equal number of unique samples ($\approx 2.04\times 10^4$) is generated on each GPU, we have set $N_s= 5n\times 10^4$ for $n$ GPUs. The green line shows the weak parallelization efficiency.}
\label{fig:weak}
\end{figure}

Fig.~\ref{fig:weak} shows the weak scaling performance of our method. We time the computation time with an increasing number of GPUs from $4$ to $64$, with each GPU calculating an approximately equal number of unique samples ($2.04\times10^4$). Our results show good weak scaling performance. Taking $32$ GPUs as a reference, we achieve a weak parallelization efficiency of approximately $93\%$ compared to $4$ GPUs.

The batched sampling algorithm we adopted takes advantage of the autoregressive properties to greatly reduce sampling overhead. But it could sacrifice parallelization due to the tree structure of our sampling process. The non-exact equivalence between the number of unique samples and the computational cost in our algorithm will generally affect both the strong and weak scaling at a larger scale when the fluctuations of the number of samples on GPUs become larger. For even larger scale parallelization in the future implementation, one could still take advantage of the conventional Monte Carlo sampling by simply implementing several independent the batch sampling algorithm, which will be effective as long as a larger number of unique samples are going to be important for that problem (the latter approach will have computational overhead if the number of unique samples is not large enough).

\section{Implications}

\begin{figure}[!htbp]
\centerline{\includegraphics[width=\columnwidth]{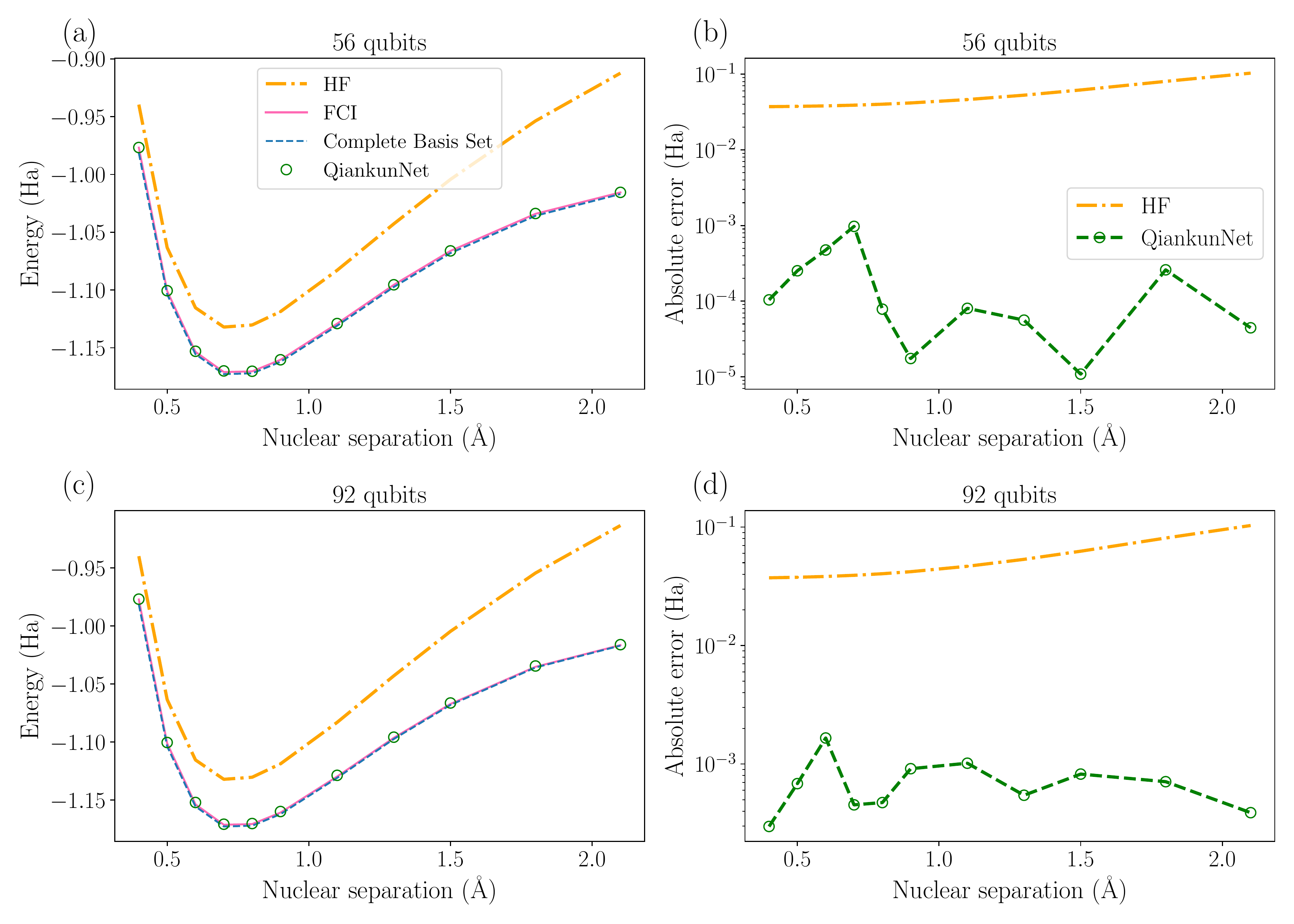}}
\caption{Potential energy surfaces of the H$_2$ molecular system in the cc-pVTZ (a) and aug-cc-pVTZ (c) basis sets calculated by our method, compared to the HF, CCSD and FCI energies. The energies calculated in the complete basis set limit (the blue dashed lines) are also plotted for comparison. The corresponding absolute errors with respect to FCI results are shown in (b,d) respectively.}
\label{fig:pes-h2}
\end{figure}

Finally, we apply our method to calculate the potential energy surfaces of the hydrogen molecular system in very large basis set: the cc-pVTZ ($56$ qubits) and aug-cc-pVTZ ($92$ qubits) basis sets. The results are shown in Fig.~\ref{fig:pes-h2}. This system is chosen because it can demonstrate the scalability of our method while being exactly solvable by FCI (in our method we have not used the knowledge that only a small number of determinants are necessary for this system, thus the computational cost for each VMC iteration is almost as difficult as for other large systems). 
For both the two large basis sets our results are in excellent agreement with the FCI results, and chemical accuracies are reached in all these simulations. We also show results obtained with FCI in the complete basis set limit, which can be considered as the ground truth for the potential energy curve of the hydrogen molecule. 
We can see that in these larger basis sets the potential energy curve already well agree the exact dissociation limit (the mean errors of our calculations with respect to the ground truth are $2.27\times 10^{-3}$ Hartree for $56$ qubits and $2.34\times 10^{-3}$ Hartree for $92$ qubits).

\section{Conclusion}

In summary, we have demonstrated a high-performance NNQS method using the transformer-based wave function ansatz for \textit{ab initio} electronic structure calculations.
To our knowledge, this is the first work which introduces transformer based architectures into NNQS to solve quantum chemistry problems. To scale up the NNQS method on distributed computing architectures, we propose a parallel batch autoregressive sampling strategy which works on multiple GPUs and a highly efficient local energy evaluation scheme which is parallelized on each GPU using multi-threading. The overall parallelization is designed to be data centric, namely each GPU manages a batch of unique samples and is responsible for all the calculations related to this batch throughout the computation, therefore minimizing data communications among different processes. 
The strong and weak scaling of our method up to $64$ GPUs demonstrate the good scalability of our method. Our work paves the way of using NNQS to study strongly correlated quantum chemistry systems with more than $100$ qubits.

Our method is mostly based on the pytorch framework and is applicable for all autoregressive wave function ansatz. It can also be easily portable to other high-performance computing architectures as long as the local energy calculation function is implemented on each process.

\begin{acks}
This work was supported by the National Natural Science Foundation of China~(Grant No. T2222026, 22003073, 11805279). The corresponding authors of this paper are Honghui Shang \\ (shanghui.ustc@gmail.com) and Chu Gu~(guochu604b@gmail.com).
\end{acks}

\bibliographystyle{unsrt}
\bibliography{./Manuscript}

\end{document}